**Reproduction Numbers $R_0$, $R_t$ for COVID-19 Infections with Gaussian Distribution of Generation Times, and of Serial Intervals including Presymptomatic Transmission**

Derek Marsh

Max-Planck Institute for Multidisciplinary Sciences[*], 37070 Göttingen, Germany

[*]Formerly, Max-Planck-Institut für biophysikalische Chemie.

E-Mail: dmarsh@gwdg.de

**Abstract**

Basic and instantaneous reproduction numbers, $R_0$ and $R_t$, are important metrics to assess progress of an epidemic and effectiveness of preventative interventions undertaken, and also to estimate coverage needed for vaccination. Reproduction numbers are related to the daily number of positive cases recorded by the national public health authorities, via the renewal equation. During periods of exponential growth or decay they are linked also to the rate constants by the Lotka-Euler equation. For either application, we need the distribution of generation times between primary and secondary infections. In practice, we use instead the directly observable serial interval between symptoms onset of infector and infectee. Pre-symptomatic transmission that occurs in COVID infection causes serial intervals to extend to negative values, which can be described with a Gaussian distribution. Consistent application of the two approaches requires careful attention to lower limits imposed on the distribution. Allowing Gaussian-distributed serial intervals to extend to minus infinity with the Lotka-Euler equation, as commonly is done, results in lower reproduction numbers than predicted from the discretized renewal equation. Here, we formulate the Lotka-Euler equation for Gaussian distributions including an explicit lower cut-off, and use this to explore the consequences of presymptomatic transmission for COVID-19 infections.

**Introduction**

Progress of the COVID-19 pandemic was quantified, in the first instance, by the number of positive cases $C_t$ recorded on day *t* by the national public health authorities. At best, cases are associated with the day that symptoms first appear, which is the closest that we get to the actual time of infection. The serial interval (SI) is the time between symptoms onset in infector and subsequent infectee (see Fig. 1, middle). We use this as a proxy for the generation time (GT) between primary and secondary infections. The probability density, $g(\tau)$, in GT relates incidence of infection, represented by daily cases, directly to the basic and instantaneous reproduction numbers, $R_0$ and $R_t$. These critical numbers characterize the severity of the disease, and its subsequent development, and also predict the extent of coverage needed for a successful vaccination campaign.

A complication arises with COVID-19 infections because infected individuals can become infectious before they develop symptoms (Du et al., 2020; Ali et al., 2020). This results in an SI distribution that extends to negative values (see Fig. 1, bottom), which can be represented by a Gaussian function but not, for instance, by the commonly used gamma or lognormal distributions. Figure 2 gives the histogram of SI-values assembled by Du et al. (2020) for identified infector-infectee pairs from the outbreak in China. This is fit best by the Gaussian distribution shown, although the histogram is not entirely symmetric. Also, unlike a Gaussian function, the histogram extends over a limited range. Particularly on the negative side, we must anticipate a cut-off, because infectiousness develops only after primary infection, and the mean incubation period is 5-6 days (Lauer et al., 2020; Bi et al., 2020; Lau et al., 2021). Comparing the bottom line with the middle line of Fig. 1, we see that the region of pre-symptomatic transmission (*pre*) cannot exceed the incubation period (*incubn*).

Here, we show how to allow for presymptomatic transmission in Gaussian distributions of SIs and GTs. Our goal is to calculate reproduction numbers from exponential rates of incidence using the Lotka-Euler equation that are consistent with those we get directly from daily case numbers by using the renewal equation. For SIs we must include a lower cut-off explicitly, and for GTs must ensure that they always stay positive.

We begin with a short mathematical background, and method of calculation. Then we compare results on COVID-19 GTs and SIs deduced by Ganyani et al. (2020), when applied to daily case data in Germany based on symptoms onset (RKI, 2025). This is followed by exploring further consequences of pre-symptomatic transmission, with two different approaches to SI-data from Ali et al. (2020) and Du et al. (2020). In the Appendix we compare the Gaussian distribution with the gamma distribution, and provide further discussion on lower limits for SIs.

There are two primary issues: (i) what are the effects of negative serial intervals that arise during pre-symptomatic transmission, and (ii) how do we truncate Gaussian distributions to ensure that generation times always stay positive

**Theoretical Background and Methods**

*Basic and Instantaneous Reproduction Numbers*, $R_0$ and $R_t$

The basic reproduction number $R_0$ is the average number of new infections produced by a typical individual throughout its infectious lifetime, when the entire population is susceptible. Expressed *per capita*, the instantaneous rate of transmission is the number per unit time $\beta(\tau)$, where $\beta(\tau).d\tau$ is the number of infections produced by an individual in time interval $\tau$ to $\tau + d\tau$ after becoming infected. The reproduction number is the sum over all $\tau$:

$$R_0 = \int_{\tau_m}^{\infty} \beta(\tau).d\tau \tag{1}$$

GTs, $\tau$, between primary and secondary infections are always positive, and the lower limit of the integral is then rigorously $\tau_m = 0$ (see Fig.1, top). However, the SI between symptoms onset of infector and infectee goes negative whenever infectiousness precedes onset of symptoms (see Fig. 1, bottom). This is the case for COVID-19 (Du et al., 2020; Ali et al., 2020). Therefore, when we use SI as proxy for GT we must retain $\tau_m$ ($\leq 0$) explicitly.

In terms of transmission rate $\beta(\tau)$, the probability density function for GT, $\tau$, between primary and secondary infections is:

$$g(\tau) = \beta(\tau)/R_0 \qquad (2)$$

where the normalizing denominator comes from Eq. 1, i.e., $g(\tau)$ is normalized over the range $\tau_m$ to $\infty$. The number of new infections, $C(t)$, at time $t$ is the sum of all infections caused by cases infected at time $\tau$ ago (i.e., at times $t - \tau$). This results in the renewal equation:

$$C(t) = \int_{\tau_m}^{\infty}[S(t)]C(t-\tau)\beta(\tau).d\tau = R_0[S(t)]\int_{\tau_m}^{\infty} C(t-\tau)g(\tau).d\tau \qquad (3)$$

where $[S(t)]$ is the fraction of the population susceptible at time $t$, and we substitute from Eq. 2 for the right-hand side. As in Eq. 1, we keep the lower integration limit general to let us replace GT by SI.

We see from Eq. 3 that the daily instantaneous reproduction number, $R_t \equiv R_0[S(t)]$, is the number of new infections $C_t$ at day $t$, divided by the total number of infective individuals causing these infections (cf. Fraser, 2007):

$$R_t = \frac{C_t}{\sum_{i=m}^{n} C_{t-\tau_i} g_{\tau_i}} \qquad (4)$$

where $\sum_{i=m}^{n} g_{\tau_i} = 1$, i.e., $g(\tau)$ is discretized over the range from day-$m$ to day-$n$ (outside this, $g(\tau) = 0$). For symmetrical distributions, such as Gaussian, $i$ is chosen symmetric about the mean. Note that, for SIs, $\tau_i$ may extend down to negative values. Then we need some values of $C_{t-\tau_i}$ for times after $t$, and thus only can get $R_t$ retrospectively. The instantaneous $R_t$ in Eq. 4 gives the number of new infections produced by an individual infected at day $t$, if conditions remain those prevailing at day $t$ (Fraser, 2007).

In regions where the rate of change in incidence varies exponentially, $C(t) = C_o \exp(rt)$, the renewal equation (Eq. 3) becomes (Wallinga and Lipsitch, 2007):

$$\frac{1}{R_0} = \int_{\tau_m}^{\infty} e^{-r\tau} g(\tau).d\tau \qquad (5)$$

The inverse of the basic reproduction number ($1/R_0$) is the Laplace transform (with lower limit $\tau_m$) of the GT probability density $g(\tau)$, with respect to the exponential rate constant $r$ for infection. This is the Lotka-Euler equation.

*Gaussian Distribution of Generation Intervals.*

The probability density of a Gaussian distribution normalized over the range $\tau_m$ to $\infty$ is:

$$g(\tau) = \frac{[1 - \Phi((\tau_m - \mu)/\sigma)]^{-1}}{\sigma\sqrt{2\pi}} \exp\left(-\frac{(\tau - \mu)^2}{2\sigma^2}\right)$$

(6)

where $\mu$ is the mean, $\sigma$ the standard deviation (SD), and $\Phi(x') = \int_{-\infty}^{x'} \exp(-\frac{1}{2}x^2).dx/\sqrt{2\pi}$ is the cumulative distribution function up to $x = x'$, for a normal distribution. (Note that when applying Eq. 4, we omit the $\Phi$-containing term from Eq. 6, because normalization is already specified by the condition: $\sum_{i=m}^{n} g_{\tau_i} = 1$.) With Eq. 6, the Laplace transform according to Eq. 5 becomes:

$$\frac{1}{R_0} = \frac{[1 - \Phi((\tau_m - \mu)/\sigma)]^{-1}}{\sigma\sqrt{2\pi}} \int_{\tau_m}^{\infty} \exp\left(-\frac{(\tau - \mu)^2 + 2\sigma^2 r\tau}{2\sigma^2}\right).d\tau$$

(7)

Rewriting the numerator in the exponential by using the identity $(\tau - \mu)^2 + 2\sigma^2 r\tau = (\tau - \mu + \sigma^2 r)^2 - (\sigma^2 r - \mu)^2 + \mu^2$, we get:

$$\frac{1}{R_0} = \frac{[1 - \Phi((\tau_m - \mu)/\sigma)]^{-1}}{\sigma\sqrt{2\pi}} \exp\left(-\mu r + \frac{1}{2}\sigma^2 r^2\right) \int_{\tau_m}^{\infty} \exp\left(-\frac{(\tau - \mu + \sigma^2 r)^2}{2\sigma^2}\right).d\tau$$

(8)

Substituting $x = (\tau - \mu + \sigma^2 r)/\sigma$, expresses the integral in standard form:

$$\frac{1}{R_0} = \frac{[1 - \Phi((\tau_m - \mu)/\sigma)]^{-1}}{\sqrt{2\pi}} \exp\left(-\mu r + \frac{1}{2}\sigma^2 r^2\right) \int_{(\tau_m - \mu)/\sigma + \sigma r}^{\infty} \exp\left(-\frac{1}{2}x^2\right).dx$$

(9)

Hence, the reproduction number becomes:

$$R_0 = \frac{1 - \Phi((\tau_m - \mu)/\sigma)}{1 - \Phi((\tau_m - \mu)/\sigma + \sigma r)} \exp\left(\mu r - \frac{1}{2}\sigma^2 r^2\right)$$

(10)

True GTs $\tau$ are always positive; this requires $g(0) = 0$ which is not part of a normal distribution. To accommodate this, consistent with Eq. 4 we choose $\tau_m \cong +1$ day, instead of zero, as the lower limit for generation times (see Fig. 2). When using instead SI as proxy, we take the explicit value of $\tau_m$ that corresponds to the lower summation limit of Eq. 4 (see Fig. 2). In principle, we could let $\tau_m \to -\infty$ for SIs, and both cumulative distribution functions in Eq. 10 then give $\Phi(-\infty) \to 0$. This yields the result commonly quoted for a Gaussian distribution:

$$R_0 = \exp\left(\mu r - \tfrac{1}{2}\sigma^2 r^2\right) \tag{11}$$

In practice, however, Eq. 11 mostly predicts values of $R_0$ that are too low. It applies only for very narrow distributions, where $\sigma \to 0$, and then it works also with generation times.

**Results**

*Comparing Generation Times with Serial Intervals*

GTs, which relate incidence of infection to reproduction numbers $R_0$ and $R_t$, are not observed directly. By using identified networks of COVID-19 infectors and infectees in Singapore and in Tianjin, Ganyani et al. (2020) estimate parameters of the GT distributions from observed SIs together with incubation times. We adopt their dataset, which associates unidentified infectors with negative SIs, because we know that pre-symptomatic transmission occurs in COVID-19 infection. Estimates of mean GTs are: 3.86 days (SD=2.65) and 2.90 days (SD=2.86) in Singapore and Tianjin (China), respectively. Correspondingly, mean SIs are similar to the mean GTs: 3.86 days (SD=4.76) and 2.90 days (SD=4.88) for Singapore and Tianjin, respectively, although the uncertainties are greater and the SDs larger (Ganyani et al., 2020).

Figure 3 shows development of the instantaneous reproduction number $R_t$ as COVID-19 infection progresses in Germany. We use incidence data based on onset of symptoms with missing data included by imputation (RKI, 2025). Although we expect onset data to be insensitive to weekend artefacts in reporting, some weekly periodicity arises from substituting uncertainly recalled onset dates by those of first medical diagnosis. We remove this anomaly with a 7-day moving average (Marsh, 2025), before applying Eq. 4 with the Gaussian distribution from Eq. 6. Lower limit of the summations is $\tau_m = +1$ day for GTs, and $\tau_m = -5$ and $-6$ days for SIs from Singapore and Tianjin, respectively. The choice of $\tau_m$ for GTs was explained already (see Fig. 2); choices for SIs correspond to $g(\tau_m)$ being reduced sufficiently close to zero (see section on lower limits in the Appendix). Solid lines use distribution parameters for Tianjin, and dotted lines those for Singapore; the latter result in somewhat higher values of $R_t$. Reproduction numbers deduced from GTs invariably are larger than those deduced from SIs. We see this particularly for the first wave of the epidemic, which illustrates the general feature that different estimates diverge more, the further $R_t$ departs from unity (Wallinga and Lipsitch, 2007).

Shapes of the overall profiles in Fig. 3 are similar because all reflect the same underlying incidence profile. Peaks in $R_t$ correspond to steepest slopes in incidence, and maxima or minima in incidence occur when $R_t = 1$. The sharp discontinuity in $R_t$ around day-107 comes from a spike in incidence during the 2020 summer trough that was associated with severe local outbreaks in the meat-processing industry and in centres of high-density housing. Small reporting anomalies appear at official holidays, e.g., Christmas, New Year and Easter, but otherwise peaks and troughs mostly follow development of the pandemic, including seasonal periodicity. Peaks at the end of the summer trough are associated with the 2020 school holidays; then follows the sharp autumn increase in incidence (days 216-233) that heralds seasonal wave-2 of the epidemic. Beyond this, further maxima in $R_t$ are associated mostly with progressive dominance of different CoV-2 variants. Emergence of Delta coincides with the 2021 school summer holidays giving the sharp rise around day-491. After this, follows a lower seasonal autumn/winter peak from day-590 onwards. Subsequent maxima in $R_t$ finally correspond to changing dominant Omicron variant lines. More details are given in Marsh (2025).

Horizontal bars in Fig. 3 come from Eq. 10, which relates $R_t$ to exponential incidence rates *r* at different stages in the epidemic, together with GT- and SI-data cited already. The bars agree well with the maxima and minima in the daily trends from Eq. 4, if we adopt values of $\tau_m$ for the lower integration limit in Eq. 10 that correspond with those given already for summations in Eq. 4. This illustrates the shortcoming of Eq. 11 for Gaussian distributions, because it results in much lower values.

*Consequences of Presymptomatic Transmission.*

Two published SI-datasets refer, in different ways, to the effects of negative SIs produced by presymptomatic transmission of SARS-Cov-2. In one, Ali et al. (2020) segment the first COVID-19 wave in mainland China outside Hubei province into pre-peak, peak, and post-peak periods. In the other, Du et al. (2020) limit a similar dataset to only positive values of SI. Estimates of mean SI by Ali et al. (2020) are: 7.8 days (SD=5.2) and 5.1 days (SD=5.0) for the pre-peak and the entire first wave, respectively. Correspondingly, mean SIs reported for the full first wave by Du et al. (2020) are: 3.96 days (SD=4.75) and 5.62 days (SD=3.92), for the complete distribution and that confined to positive values $\tau > 0$, respectively.

Figure 4 shows development of $R_t$ in Germany, based on the same incidence data as in Fig. 3, but using the SI-data sets just described. As expected, the shape of the profiles is similar to that in Fig. 3, but numerical values differ. Horizontal bars again relate $R_t$ to exponential incidence rates *r* in the same incidence timeline. Lower limits of the summations and integrals (Eqs. 4 and 10) are $\tau_m = -2, -5$ and $+1$ days for the pre-peak, full peak and only positive SIs, respectively. Black lines in Fig. 4 are for the pre-peak, and grey lines cover the full peak. Values of $R_t$ for the pre-peak period are considerably higher than those for the full peak because the mean SI shortens as the pandemic proceeds, which results from growing public awareness and preventative official interventions (Ali et al., 2020). Light grey lines in Fig. 4 are for the complete range of SIs, and dotted lines only for positive SIs (Du et al., 2020). When negative values of SI are excluded, a Gaussian function no longer best describes the truncated dataset. The authors caution against placing undue reliance on data truncation, but find that a lognormal distribution (see Appendix, Eq. A.1) fits the $\tau > 0$ regime best (Du et al., 2020). The resulting profile (dashed line) lies close to that based on the Gaussian distribution of SIs, which includes the negative values (light-grey solid line).

**Discussion**.

The principal message here is that, when allowing for pre-symptomatic transmission based on a Gaussian distribution of SIs, Eq. 10 with $\tau_m$ approximately equal to the incubation period (−5 to −6 days for COVID), and not Eq. 11, should be used for determining the reproduction numbers $R_0$ and $R_t$. The section of the Appendix on lower limits for negative SIs explains the reasoning behind this choice. For Gaussian distributions of GTs, taking $\tau_m = +1$ day ensures that only positive GTs are allowed (see Fig. 2).

Truncation is not confined solely to situations with pre-symptomatic transmission. Quite generally, we must decide in how many points to discretize the SI- or GI-distribution when calculating instantaneous $R_t$s for the summation in Eq. 4. Reference to the histogram in Fig. 2 shows us that the actual number of data points is limited and does not extend as far out as their representation by continuous functions, especially Gaussians. For GIs, a Gaussian without lower cut-off is not the most appropriate distribution, unless all appreciable probabilities are confined solely to positive GIs. This may occur for small SDs, i.e., sharp distributions. Otherwise, distributions that extend only from zero upwards, such as lognormal or gamma distributions (Eqs. A.1, A.2), may provide better fits.

When fitting COVID SI-data that does not include negative values from pre-symptomatic transmission, a gamma distribution gives the best fit in several cases (Cereda et al., 2020; Bi et al., 2020; Li et al., 2020). Unlike Gaussians, gamma distributions always remain positive and potentially could fit GI-data directly without truncation. In Fig. A.1 of the Appendix, we compare the $R_t$-profiles deduced for Tianjin and Singapore in Fig. 3 using Gaussian GT-distributions with those deduced assuming gamma distributions having the same means and SDs. The values of $R_t$ for gamma distributions lie below the corresponding ones for Gaussian distributions (in the sense of excursions, both positive and negative, from $R_t = 1$). Nevertheless, the values remain somewhat comparable, especially at later times when $R_t$ is closer to unity. Basic reproduction numbers are the highest, with values from Singapore GT-data of $R_0 = 2.8$ and 2.3 for Gaussian and gamma GT-distributions, respectively. Note that $R_t$-values from the gamma GT-distributions remain wholly above those from the Gaussian *SI-distributions* given in Fig. 3, and even further above those predicted by Eq. 11 that often is used for Gaussian distributions.

**Acknowledgement**

I gratefully thank Christian Griesinger and the Department of NMR-based Structural Biology for essential support.

**Figure legends**

Fig.1. Line of infection (infect$_i$). Top: for generation time (*GT*); middle: for incubation time (*incubn*) before onset of symptoms (onset$_i$) with serial interval (*SI*); bottom: for the case of pre-symptomatic transmission (*pre*).

Fig. 2. Serial interval distribution (468 infections; histogram) and Gaussian fit (mean=3.96 day, SD=4.75; circles) for mainland China (Du et al., 2020). Dashed lines illustrate truncation of the Gaussian at $\tau_m$, for serial interval (SI) and for generation time (GT).

Fig. 3. Instantaneous reproduction numbers R$_t$, based on GTs and SIs (as indicated), for COVID-19 infection over the years 2020-2022 in Germany. Deduced from 7-day averaged incidences, based on onset of symptoms (Eqs. 4 and 6, with daily incidences from RKI, 2025). Horizontal bars are values of R$_t$ ($\equiv R_0(t)$) for Gaussian distributions of GT and SI (Eq. 10); means and SDs of the distribution for Singapore (grey) and Tianjin (black) are from Ganyani et al. (2020); exponential incidence rates *r* are from Marsh (2025). *y*-axis is logarithmic.

Fig. 4. COVID-19 reproduction numbers R$_t$, based on different Gaussian distributions of SIs (as indicated), over years 2020-2022 in Germany. Derived as in Fig. 3. Dotted line is exceptionally for a lognormal distribution, which allows only positive SIs (Du et al., 2020). Horizontal bars are R$_t$ deduced from exponential incidence rates *r* (Eq. 10); means and SDs of the distribution for the pre-peak (black) and all (grey) of the first COVID-19 wave in mainland China outside Hubei province from Ali et al. (2020), and (in light grey) from Du et al. (2020). *y*-axis is logarithmic.

Fig. A.1. Comparison of R$_t$-profiles for Germany, deduced from Gaussian (black lines) and gamma (light grey lines) distributions of GTs, derived as in Fig. 3. Solid lines are from means and SDs of the GT-distribution in Tianjin, and dotted lines are for Singapore (Ganyani et al., 2020). Horizontal bars are R$_t$ from exponential incidence rates *r* (Eq. 5); Gaussian distributions in Tianjin (black), Singapore (grey); gamma distributions in Tianjin (solid light grey), Singapore (dotted light grey).

Fig. 1.

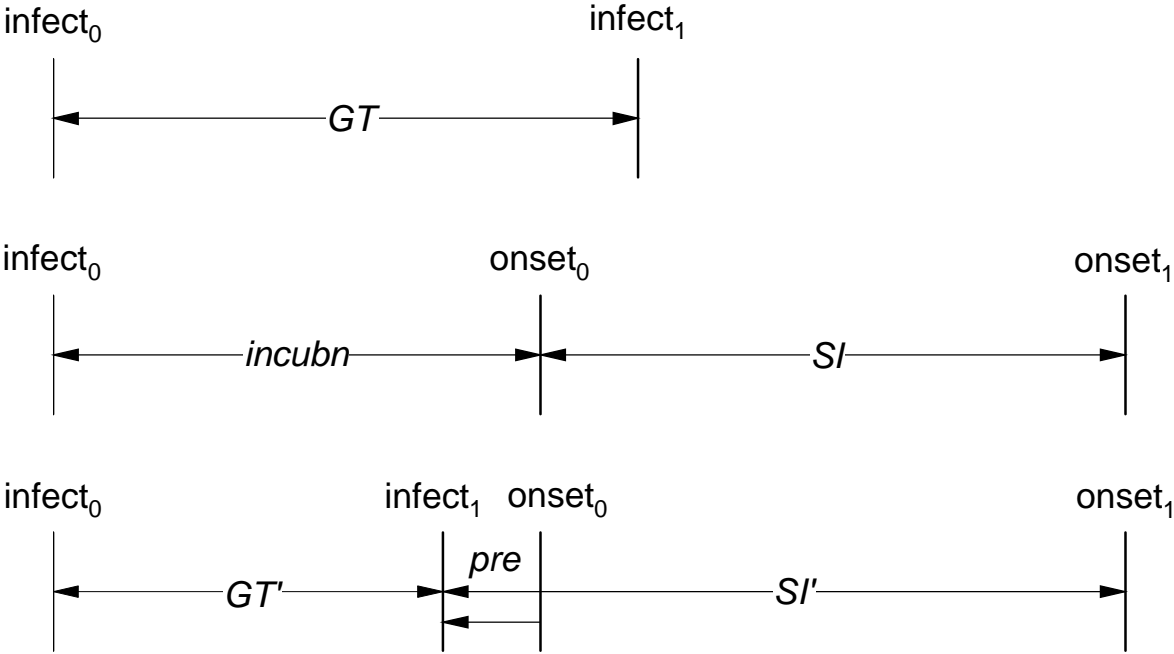

Fig. 2.

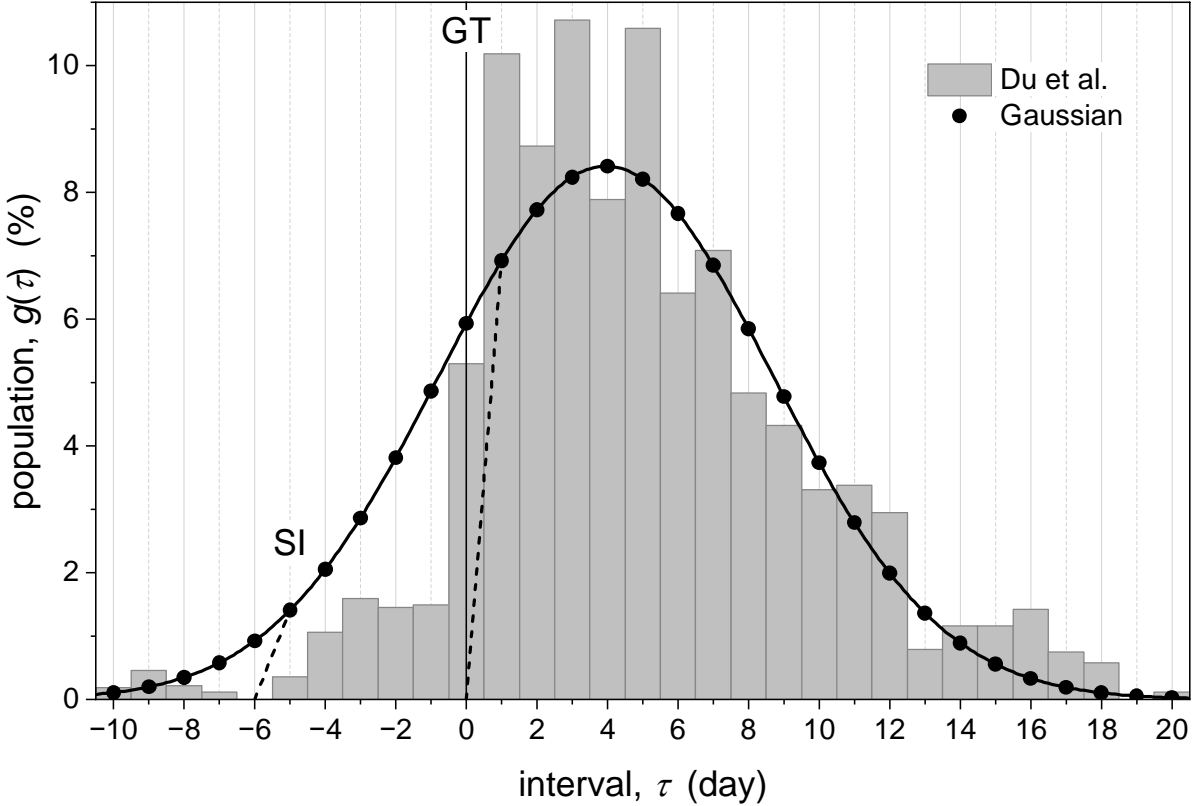

Fig. 3.

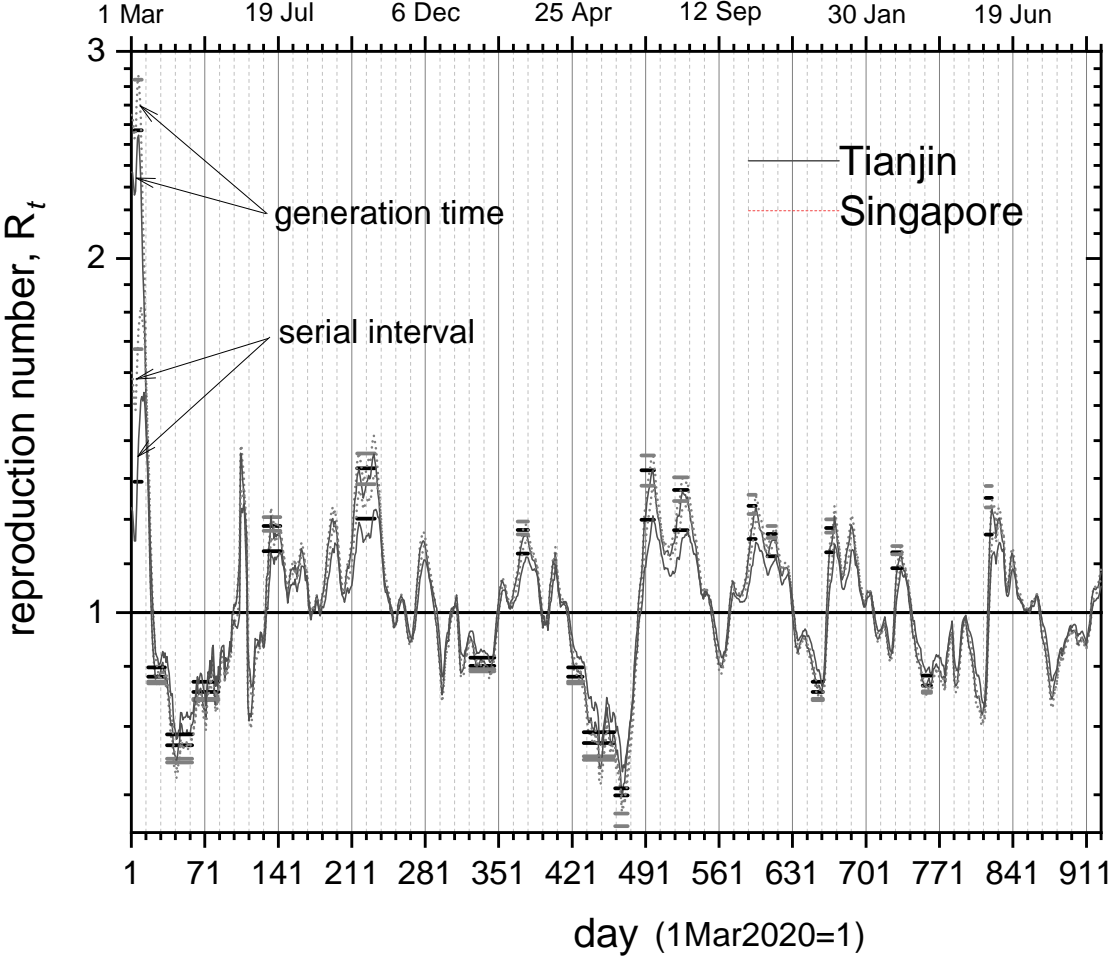

Fig. 4.

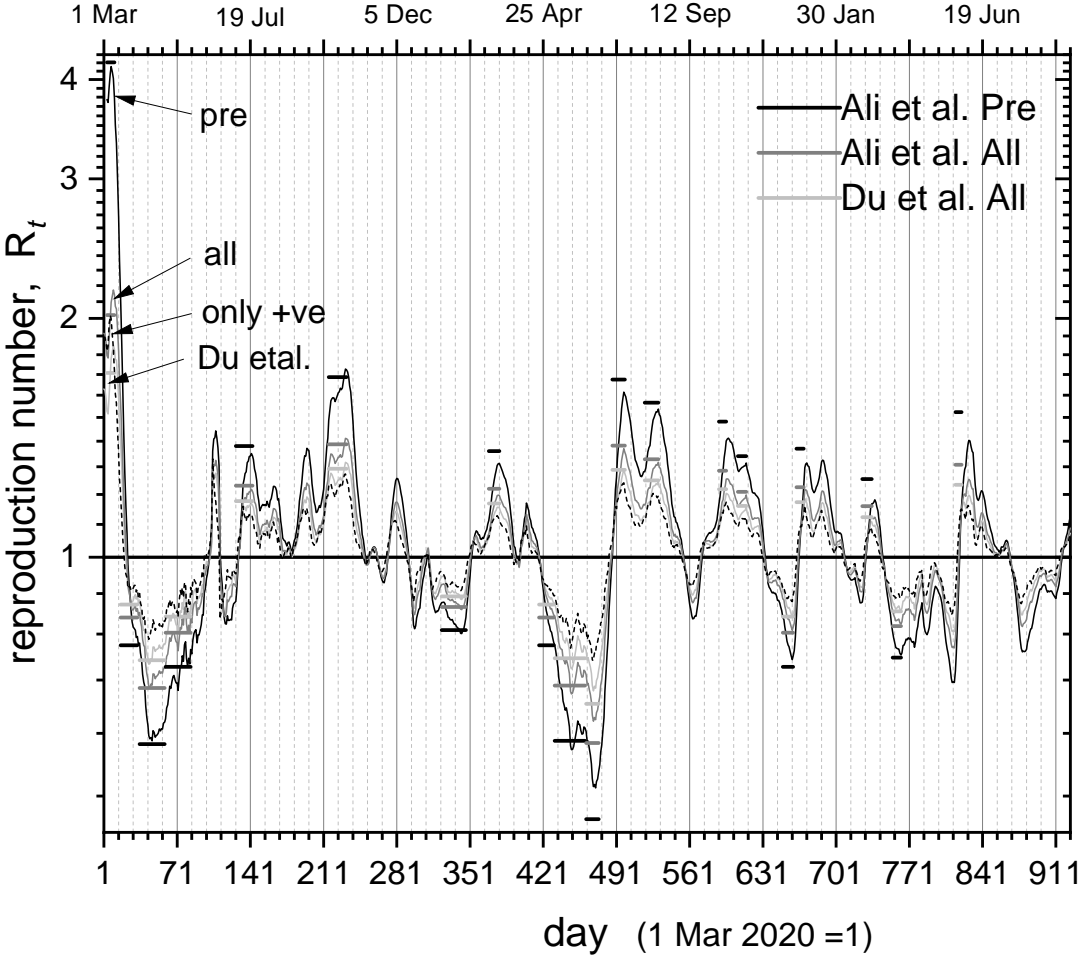

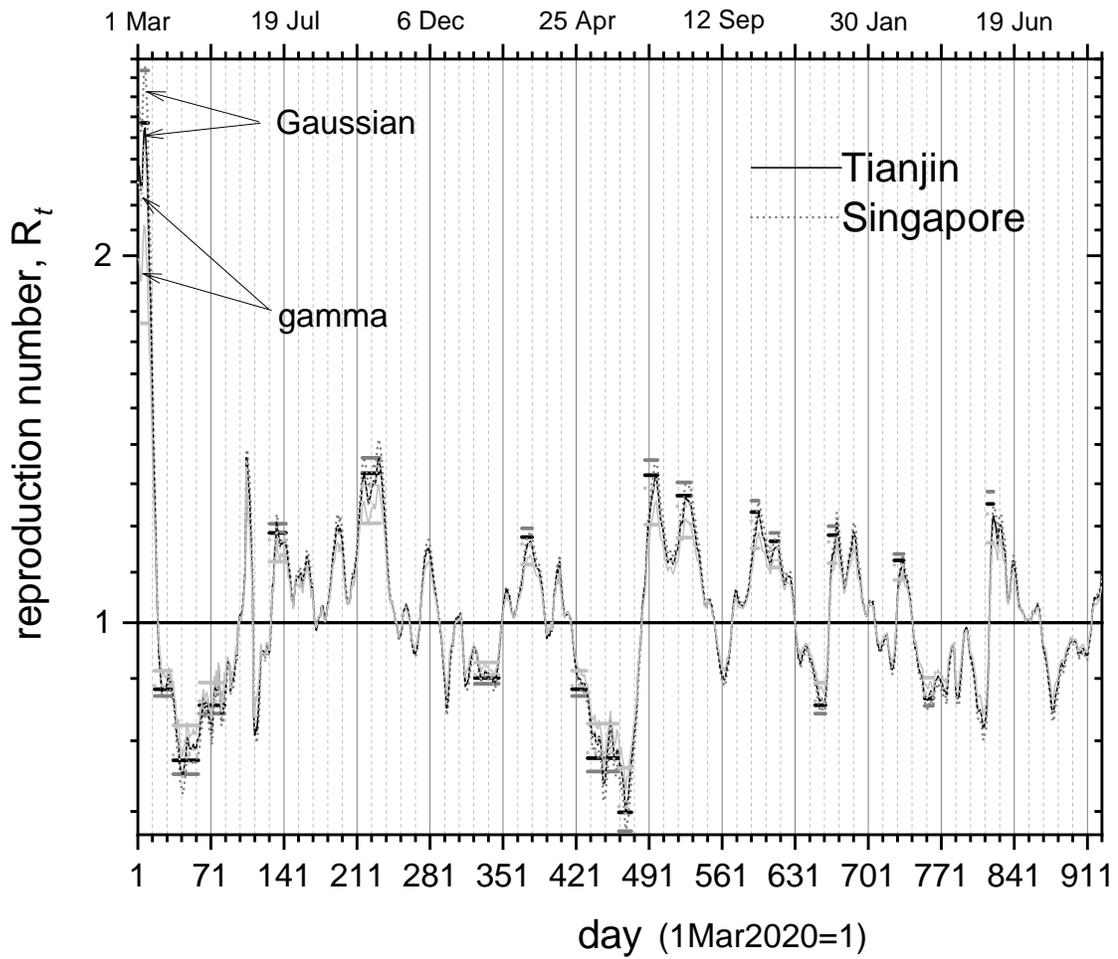

Fig. A.1.

# Appendix

*Lognormal Distribution of Serial Intervals.*

The probability density function for a lognormal distribution of SIs, $\tau$, is:

$$g(\tau) = \frac{1}{\tau\sigma\sqrt{2\pi}}\exp\left(-\frac{(\ln\tau - \mu)^2}{2\sigma^2}\right)$$

(A.1)

where the mean is $\exp(\mu + \frac{1}{2}\sigma^2)$, and the variance is $(SD)^2 = [\exp(\sigma^2) - 1]\exp(2\mu + \sigma^2)$. This distribution is non-zero only for positive values of $\tau$, and $g(0) = 0$. For the truncated SI-data of Du et al. (2020), best fitting parameters are: $\mu = \ln(2.02)$ and $\sigma = \ln(2.78)$.

*Gamma Distribution of Generation Times.*

The probability density function for a gamma distribution of generation times, $\tau$, is:

$$g(\tau) = \frac{\lambda^\alpha}{\Gamma(\alpha)}\tau^{\alpha-1}\exp(-\lambda\tau)$$

(A.2)

where $\Gamma(\alpha)$ is the gamma function. The mean is $\mu = \alpha/\lambda$, and the standard deviation is $SD = \sqrt{\alpha}/\lambda$. This distribution is non-zero only for positive values of $\tau$, and $g(0) = 0$. From Eq. 5 with $\tau_m = 0$, the reproduction number is:

$$R_0 = (1 + r/\lambda)^\alpha$$

(A.3)

where we get $\alpha$ and $\lambda$ from the mean and SD.

Fig. A.1 compares the $R_t$-profiles deduced for Gaussian GT-distributions (black lines) taken from Fig. 3 with those deduced using gamma distributions (light grey lines) having the same means and SDs, from both Tianjin (solid lines) and Singapore (dotted lines). Shapes of the profiles are very similar for the two distributions because they derive from the same incidence timeline. Horizontal bars are $R_t$ from exponential incidence rates $r$, using Eq. A.3 for the case of the gamma distribution; black and grey bars are for Gaussian distributions in Tianjin and Singapore, respectively, and solid and dotted light grey bars are similarly from gamma distributions.

For both Gaussian and gamma distributions, $R_t$-values are somewhat higher using Singapore GTs than those using Tianjin GTs. The values of $R_t$ for gamma distributions are lower than the corresponding ones for Gaussian distributions. (In fact, the *gamma* $R_t$-profile for Singapore in Fig. A.1 superimposes almost completely on the *Gaussian* $R_t$-profile for Tianjin.) Nonetheless, values of $R_t$ from gamma GT-distributions remain wholly above those from Gaussian *SI-distributions* given in Fig. 3, and even further above those predicted by Eq. 11 for Gaussian distributions. For instance, with Singapore data, basic reproduction numbers for Gaussian and gamma GT-distributions, Gaussian SI-distributions, and Eq. 11 are: $R_0 = 2.84, 2.27, 1.67$ and 1.29, respectively. With Tianjin data, corresponding values are: 2.57, 1.76, 1.29 and 0.95, respectively (see Figs. 3 and A.1).

*Including upper integration limit, $\tau_{up}$.*

Although less critical, we include for completeness an explicit upper limit $\tau_{up}$ in the integrals. Eq. 10 then becomes:

$$R_0 = \frac{\Phi((\tau_{up} - \mu)/\sigma) - \Phi((\tau_m - \mu)/\sigma)}{\Phi((\tau_{up} - \mu)/\sigma + \sigma r) - \Phi((\tau_m - \mu)/\sigma + \sigma r)} \exp\left(\mu r - \tfrac{1}{2}\sigma^2 r^2\right)$$

(A.5)

where $\tau_{up} \to \infty$ leads to $\Phi(\cdot) \to 1$, and we recover Eq. 10. Decreasing $\tau_{up}$ reduces $R_0$, as expected, but not greatly for realistic cases. For instance, Fig. 3 gives $R_0 = 1.67$ and 1.29 deduced from Eq. 10 with SI-data for Singapore and Tianjin, respectively, and this reduces to $R_0 = 1.63$ and 1.25 when using Eq. A.5 with $\tau_{up} = 13$ days (or 12 days for Tianjin). Increasing $\tau_{up}$ to 17 days (or 16 days) brings the total number of SI-data points up to 25, and already restores $R_0$ close to the original values obtained from Eq. 10.

*Lower Limits for Negative Serial Intervals.*

When SIs can go negative, left censoring is a significant issue. We can alleviate this somewhat by extrapolating the initial exponential dependence of COVID incidence to earlier times. When discretizing symmetric SI-distributions, such as Gaussian, it is convenient to choose symmetrically located pairs of data points. However, to obtain sufficient coverage of the initial parts of the incidence timeline in Eq. 4, we need to restrict the range of negative $\tau_i$s. As noted already, this range is limited physically by the length of the incubation time (see Fig. 1). Then, to allow realistic and consistent comparisons, we must adopt the same lower limit, $\tau_m$, in both Eqs. 4 and 5.

As an empirical approach, we choose here that $g(\tau_m) \leq 0.01 - 0.02$. This results in discretization over a range of 21-23 points, including outer zeroes, which we can increase by adding further positive SIs. Typically, histograms of real data that include negative SIs contain ca. 30 bars (see Fig. 2 and Du et al., 2020; Ali et al., 2020). In the previous section, we noted that increasing $\tau_{up}$ to give a total of 25 points already comes close to full coverage. (Originally limiting to symmetrically placed pairs yields a total of 21 points.)

The criterion that $g(\tau_m) \leq 0.01 - 0.02$ typically results here in $\tau_m = -5$ to $-6$ days, which as noted already is close to the mean incubation time of SARS-Cov-2. If we extend the lower limit in Eq. 10 by a further day to $\tau_m = -6$ and $-7$ days for Singapore and Tianjin, respectively, the basic reproduction numbers go down from $R_0 = 1.67$ and 1.29, respectively (see horizontal bars in Fig. 3), to $R_0 = 1.56$ and 1.20. This is still an appreciable effect, but Fig. 1 suggests that cut-offs no longer than the mean incubation time are a physically reasonable choice. Examining the SI-histogram for mainland China in Fig. 2 also gives practical support for choosing $\tau_m = -5$ to $-6$ days as an appropriate cut-off for COVID-19.